\documentclass{article}

\usepackage[margin=1.4in]{geometry}
\usepackage{fancyvrb}
\usepackage{enumitem}

\usepackage{tikz}

\newlength{\RoundedBoxWidth}
\newsavebox{\GrayRoundedBox}
\newenvironment{prolbox}[1][\dimexpr\textwidth-1.5ex]%
   {\setlength{\RoundedBoxWidth}{\dimexpr#1}
    \begin{lrbox}{\GrayRoundedBox}
       \begin{minipage}{\RoundedBoxWidth}}%
   {   \end{minipage}
    \end{lrbox}
    \begin{center}
    \begin{tikzpicture}%
       \draw node[draw=yellow!16,fill=yellow!16,%
             inner sep=0ex,text width=\RoundedBoxWidth]%
             {\usebox{\GrayRoundedBox}};
    \end{tikzpicture}
    \end{center}}

\usepackage{marginnote}
\begin{document}

\def\myname{Ond\v{r}ej B\'ilka}
\def \floor#1{\lfloor #1 \rfloor}
\def \ceil#1{\lceil #1 \rceil }
\def\dir{\mathop{\rm dir}}
\def\mod{\mathop{\rm mod}}
\def\tg{\mathop{\rm tg}}

\def\regreg{{$\mathrm{REG^{REG}}$}}
\def\peg{$\mathrm{PEG}$}
\def\extext#1{{\bf\it ``#1''}}

\input epsf
\newtheorem{thm}[section]{Theorem}
\newtheorem{qst}[section]{Question}
\newtheorem{obs}[section]{Observation}
\newtheorem{cor}[section]{Corollary}
\newtheorem{claim}[section]{Claim}
\newtheorem{lem}[section]{Lemma}
\centerline{\Large Structured Grammars are Effective}

\centerline{ \myname}
\centerline{ Department of Applied Mathematics}
\centerline{ Charles University}
\centerline{ Malostransk\'e N\'am. 25}
\centerline{ 118 00 Praha 1, Czech Republic}
\centerline{ ondra@kam.mff.cuni.cz}

\begin{abstract}
Top-down parsing has received much attention recently. Parsing expression grammars (\peg{}) allows construction of linear time parsers using packrat algorithm.
These techniques however suffer from problem of prefix hiding. We use alternative formalism of relativized regular expressions \regreg{} for which top-down backtracking parser runs in linear time. This formalism allows to construct fast parsers with modest memory requirements for practical grammars. We show that our formalism is equivalent to \peg{}.
\end{abstract}
\section{Introduction}
A top-down parsing implementation can be viewed as bunch of mutually recursive functions recognizing individual rules in grammar description. Naively implemented parser of the following rule:\\
\verb$R="aa" R | "a" R$ \\
on \extext{aaaaaaaaaaaaaaaaaaa...} can take exponential time. 

Incorporating left recursion also causes problems. A naive parser of \\
\verb$L=L a$ \\
would call L infinitely many times. Various approaches were suggested to solve both problems.

In natural language processing we typically want to enumerate possible interpretations of ambiguous grammar.

Frost \cite{frost} gave $O(n^4)$ algorithm that outputs compact representation of all parses \cite{tomita} and handles left recursion as recursive descend. 
Parsing expression grammars allow unlimited lookahead. Okhotin \cite{bool} suggest to extend context free grammars with lookahead to class of {\it boolean grammars}. Again his algorithm for boolean grammars had complexity $O(n^4)$. Both these algorithms were improved by variant of Valiant algorithm \cite{valiant} to obtain complexity $O(M(n) \log n)$ where $M(n)$ is time of matrix multiplication. When boolean grammars are restricted to {\it unambiguous boolean grammars} there exists $O(n^2)$ algorithm. 

For programming languages ambiguity is undesirable. One of approaches are parsing expression grammars defined by Ford \cite{ford}. A parsing expression grammars (\peg{} for short) can be viewed as a top-down parser that places three additional constraints. First is that rules are deterministic. Second is restricting choice operator \verb$|$ to {\it ordered choice} operator \verb$/$. Once an alternative of ordered choice succeeds then choice succeeds and we do not backtrack if something fails later. Third is that iteration is greedy and does not backtrack. 

This definition without backtracking introduced problem of {\it prefix hiding}, an expression \\
\verb$"a"/"ab"$    does not match string \extext{ab}.

Seaton in his Katahdin language \cite{katahdin} uses different {\it longest choice} operator to partially solve this problem. A longest choice tries all alternatives and deterministically chooses a longest match. However this does not eliminate the prefix hiding completely. Parser of:\\
\texttt{"\textvisiblespace"* "\textvisiblespace{}foo"} (\verb$*$ is iteration operator) still does not match \extext{\textvisiblespace{}foo}. 

We take another approach. Programming languages use only two types of recursion: iteration and nested recursion. By making this information explicit we can generate linear time parsers that are equivalent to the fully backtracking ones.

We present new formalism of relativized regular expressions  \regreg{}. Our formalism relaxes determinism of \peg{} grammars. As in \peg{} we support arbitrary lookaheads. Previous results can be easily derived using  \regreg{} formalism.

Although \regreg{} seems stronger than  \peg{} we show that \peg{} and  \regreg{} are equivalent.

Author's Amethyst language implements \regreg{} parser with several extensions. One is that Amethyst allows parametrized rules like \verb$times(4,'foo')$ and support lambdas like \verb$times(3,(| a|b|c |))$. Amethyst language is described in author's thesis \cite{thesis}.

\section{Structured grammars}

We devise an approach to describe programming languages which we call {\it structured grammars}. We build on an analogy with structured programming languages. 

As programs used arbitrary {\it goto} constructs, grammars use arbitrary forms of recursion.
To make programs more readable, programming languages was extended by adding structured control flow constructs
making it easier for developers to read the code on a local basis without spending hours to understand
the whole context. We seek similar goals with introduction of structured grammars.

Assume we are given a grammar for the fully-backtracking top-down parser. We say it is {\it structured grammar} if it satisfies following conditions:

\begin{enumerate}[label*=\arabic*.]
\item {\it Transparency of semantic actions}. We can imagine that parser is augmented by an oracle that may decide that alternative will eventually fail. The parser should display same output regardless if we tried that alternative and failed or used the hint from the oracle. Lookaheads form important case. We always revert actions made by lookaheads.

\item Recursion is restricted to {\it iterative} and {\it nested recursion}.\\
Iterative:\\
 For example arguments of function in C are lists of \verb$expressions$ separated by "\verb$,$". We typically use iteration \verb$*$ operator. Iteration can be also described by left recursive or by right recursive rules. When possible iteration should be written in way that is associative.\\
Nested recursion:\\
What is not iteration can be described by {\it \texttt{start}} and {\it \texttt{end}} delimiters.  
We require user to annotate this concept by operator \verb$nested(${\it \texttt{start,middle,end}}\verb$)$. 

Simplest example are properly parenthised expressions. They can be described as:\\
\verb$exp = nested('(',(| exp |),')')$\\
We show two less trivial examples in Amethyst language syntax. A while loop in \verb$C$ is matched by:\\
\verb$'while' exp nested('{',(| stmts |),'}')$ \\
Python uses indentation to describe nesting. We use a semantic predicate to find where we end. We match python while loops in python as:\\
\verb$ nested((|   '\n' ' '*:x 'while' exp |),(| stmts |),$\\
\verb$        (| &('\n' ' '*:y &{x.size>y.size}) |))$

A nesting should satisfy three natural conditions.
\begin{enumerate}[label*=\arabic*.]
\item Position of {\it\texttt{end}} delimiter is determined by position {\it\texttt{start}} delimiter.
\item When \verb$nested$ starts in smaller position it should end in strictly larger position.
\item When both \verb$nested(${\it \texttt{start,mid1,end}}\verb$)$ and  \verb$nested(${\it \texttt{start,mid2,end}}\verb$)$ match string then their end positions should agree.
\end{enumerate}
\end{enumerate}

Note that programming languages implicitly follow this convention. Other types of recursion are undesirable because user cannot reason about them locally.

One of reasons is that programming languages were described as deterministic context free grammars. Thus they can be written by deterministic push-down automaton. We can model push/pop pair by calling \verb$nested$. Indeed if we did not include lookaheads our class would be equivalent to class of deterministic context free grammars. We leave proof as an exercise.

Structured grammars offer additional advantages. 
For example we can use the structure information to semiautomatically construct error correction tool.

For equivalence with top-down parser our parsing algorithm needs condition 2.1. Without condition 2.2 a parser would be quadratic instead linear time. Condition 2.3 is design guideline which is not needed in our algorithm.

\section{\peg{} and \regreg{} operators.} 

A parsing expression grammars \cite{ford} are defined by the following operators.

\noindent\begin{tabular}{| l | l | }
\hline
\verb$'s'     $& Match string.\\
\verb$ r      $& Rule application.\\
\verb$e1 e2   $& Sequencing.\\
\verb$e1/e2   $& Ordered choice.\\
\verb$ e*   e+$& Iteration.\\
\verb$&e   ~e $& Positive and negative lookahead.\\
\verb${a} &{a}$& Semantic action and predicate.\\
\hline
\end{tabular}

We relax determinism of \peg{} to \regreg{} expressions. We can describe every structured grammar by \regreg{} rules with linear time guarantee.

A \regreg{} expressions mostly use the same operators as \peg{}. Difference is that operators do backtracking except of \verb$nested$ which behaves deterministically. 

\noindent\begin{tabular}{| l | l |}
\hline
\verb$nested(start,mid,end)$& Nested operator.\\
\verb$e1|e2 $& Priorized choice.\\
\verb$e* e+ $& Backtracking iteration.\\
\verb$e1[e2]$& Enter operator.\\
\hline
\end{tabular}

Enter operator is described in section \ref{semact}.
\subsection{Simple algorithm}\label{simplealg}
We will describe our parser in functional programming style pseudocode. We denote lambda as: \\
\verb$\lambda(arguments){body}$ and call it with \verb$call$ method.\\
Rest of code is self explanatory.

We start with simple implementation and will progressively add more details. 

A \regreg{} parser behaves mostly as a top-down parser.  We use function \verb$match(e,s,cont)$ where \verb$e$ is expression we match, \verb$s$ is current position and \verb$cont$ is continuation represented as lambda.
\begin{prolbox}
\begin{verbatim}
match( r  ,s,cont) = match(body(r),s,cont)

match('c' ,s,cont) = if s.head=='c'              ;cont.call(s.tail)
                     else                        ;fail

match(e f ,s,cont) = match(e,s,\(s2){ match(f,s2,cont) }

match(e|f ,s,cont) = if match(e,s,cont)          ;success
                     else                        ;match(f,s,cont)

match(~e  ,s,cont) = if match(e,s,\(s2){success});fail
                     else                        ;cont.call(s)

match(e*  ,s,cont) =
  cont2 <-    \(s2){ if match(e,s2,cont2)        ;success
                     else                        ;cont.call(s2) 
                   }
  cont2.call
\end{verbatim}
\end{prolbox}
Pseudocode above describe naive top-down parser. For \regreg{} class we restrict recursion and add \verb$nested$ operator:
\begin{prolbox}
\begin{verbatim}
match(nested(st,mi,en) ,s,cont) =
  s3 <- match((st mi en),s,\(s2){success})
  if s3    ; cont.call(s3)
  else     ; fail
\end{verbatim}
\end{prolbox}

\subsection{Equivalence with top-down parsers and \peg{}}

We prove that for structured grammars  \regreg{} parser finds same derivation as fully backtracking one. As top-down parser does not directly support left recursion we do not consider left recursion in this section. 

An implementation of the fully backtracking parser is same as the implementation of \regreg{} parser in section \ref{simplealg} except of \verb$nested$:

\begin{prolbox}
\begin{verbatim}
match(nested(st,mi,en), s, cont) = match((st mi en), s, cont)
\end{verbatim}
\end{prolbox}

For sake of proof we transform rewrite implementation of \verb$nested$ in \regreg parser to equivalent one. In \verb$nested$ we only consider first alternative in the way following pseudocode suggests:
\begin{prolbox}
\begin{verbatim}
match(nested(st,mi,en), s, cont) = first <- true
  match(s, (st mi en), \(s2){
    if first ; first <- false
             ; cont.call(s2)
    else     ; fail 
  })
\end{verbatim}
\end{prolbox}

An equivalence with top-down parser can be proved by easy induction on the nesting level. 
\begin{enumerate}
\item When expression contain no nesting we have identical implementation. 

\item Assume we proved proposition for nesting level $\ell-1$. We prove level $\ell$ by second induction on the number of \verb$nested$ calls in the continuation on level $\ell-1$.
\begin{enumerate}
\item For continuation that does not call \verb$nested$ we use same argument as in 1.

\item Assume we have continuation that calls \verb$nested$ $n$ times. Consider first time we call \verb$nested$. If this call fails it, by induction, also fails in the fully backtracking parser and we are done.

Otherwise  \regreg{} and the fully backtracking parser first try lexicographically smallest alternative in the recursion tree. If a continuation succeeds a derivation is same by induction.

If a continuation fails we use assumption 2.1. of structured grammars.
Our parser does not try alternatives further. A backtracking parser enumerates all derivations. As every derivation ends in same position and continuation will always fail. Thus the backtracking parser behaves like \regreg{} parser.
\end{enumerate}

\end{enumerate}

Like not every C program is structured program not every  \regreg{} grammar is structured one. We can use \verb$nested$ with empty {\it\texttt{start}} and {\it\texttt{end}} to implement \peg{} operators. This gives us inclusion \peg{} $\subseteq$ \regreg{}. An opposite inclusion is true but not very enlightening. As there are only finitely many pairs $(e,cont)$ we can for each pair write a \peg{} rule that emulates  \regreg{} algorithm.

For linear time guarantee we still require every recursion except left and right recursion to be annotated by \verb$nested$.

\section{Relativized regular machines. }
To better understand languages recognized by relativized regular expressions we introduce the {\it relativized regular machines} that are similar to nondeterministic finite state machines \cite{nfa}. We use this formalism as an inspiration for effective low-level implementation of parsers.

It is easy to see that an continuation corresponds to right congruence class. We use representation that unifies identical expressions and continuations. This can be viewed as NFA state minimization\footnote{NFA minimization is NP-hard in general case. Our approach is a good heuristic.}. 

A {\it relativized regular machine} is similar to nondeterministic finite state machine. A machine can be described by triple 
$M=(S,t,a)$ where 

\begin{tabular}{l}
S is set of states, \\
$t: (S,N,S) \to (M,S)$ set of transitions and \\
$a \subset S$ a set of accepting states.\\
\end{tabular}

We have elementary machines that match single character. 

Transitions from state $s$ are done in following way. We put $(M_i,s_i)=t((s,i,r_{i-1}))$ then recursively call machine $M_i$ and if it succeeds we move to its end position and set state to $s_i$. Based of accepting state $r_i$ this choice reaches we choose a next choice. 

\section{Effective implementation} \label{implement}
A implementation above runs in linear time but constant factor is quite high. For better constant factor our parser generator applies various optimizations. We use a low-level representation that is suitable for these optimizations.

In this section we describe parser that does not consider semantic actions. Semantic actions will be added in next section. 

Representation of expressions is similar to syntax tree but more compact. We use similar technique as compact representation of derivations  in Tomita algorithm \cite{tomita}:

\begin{enumerate}
\item All nodes are immutable.

\item We represent all identical subtrees by single object. When we are asked to construct a node optimizer first tries to simplify node by algebraic identities. If after simplification we obtain node identical to previously constructed node we return previously constructed node.
\end{enumerate}

We will again use function \verb$match(e,Args[ ... ] ) -> Result[ ... ]$.

We will extend several times what \verb$Args$ and \verb$Result$ objects contain. Initially we define following fields:

\begin{tabular}{l l}
\verb$Args.s$ &is starting position of string,\\
\verb$Result.s$  &is end position of string,  \\
\verb$Args.cont$ &is a continuation.\\
\end{tabular}

Objects \verb$Args$ and \verb$Result$ have method \verb$change$ that creates new object with appropriately changed fields.

\subsection{Sequencing}
We represent sequencing operator \verb$head tail$ by object with pattern \verb$Seq[head tail]$. Representing sequencing in this way allows \verb$tail$ parts to be shared.
Implementation is straightforward.

\begin{prolbox}
\begin{verbatim}
match( Seq[head tail],a) = match(head,a.change(
  cont:\(a2){
    match(tail,a.change(cont: a.cont))
  }))
\end{verbatim}
\end{prolbox}

\subsection{Choice and lookaheads}

Inspired by relativized regular machines we model choice and lookaheads by more general \verb$Switch$ operator. First we need add field \verb$Result.state$. This state will be used to pass information from rules to the \verb$Switch$ operator.

A \verb$Switch$ operator satisfies following pattern \verb$Switch[ head alt:{state=>tail} merge ]$.
\verb$Switch$ operator first matches a head. Then it looks what end state head reached and matches tail entry corresponding to that state. Finally it computes final state from states of children by merge method. 

For simplicity in this paper we use only two states  \verb$success$ and \verb$fail$. We use identity function as a merge method. We also add \verb$success$ and \verb$fail$ rules with obvious implementation:
\begin{verbatim}
match(Rule[success]) = Result[state: success]
match(Rule[fail   ]) = Result[state: fail   ]
\end{verbatim}

This is quite general operator and we illustrate its uses on several examples.

A choice operator backtracks until success state was reached. An implementation is:

\begin{verbatim}
e1|e2  ->    Switch[ e1 {success: success
                            fail: e2}]
\end{verbatim}

Lookaheads can be modeled like:\\
\begin{verbatim}
~e   ->    Switch[ Seq[e success ]
                  {success:   fail,
                   fail:      empty} ]
&e   ->    Switch[ Seq[e success ]
                  {success:   empty,
                   fail:      fail } ]
\end{verbatim}

A \verb$Switch$  makes optimizations easy.  

Switches can be easily composed. To compose switches \verb$A$ and \verb$B$ a simplest way is to use states that are pairs (state from \verb$A$,state from \verb$B$). We need to define merge method to compute final state. We can represent these pairs compactly as bit vector.

Another optimization is predication. When we know first character we can simplify expression:

\begin{verbatim}
Switch[ Result[ first_character ]
                  { 'a': expressions that can start by a,
                    'b': expressions that can start by b, 
                    ...   
\end{verbatim}

For choice \verb$e1|e2$ we can, based on result of the partial match of e1, simplify matching of \verb$e2$. For example consider expression:

\begin{verbatim}
  (a|b) c (d|f)
| (b|c) c f
\end{verbatim}
on string \extext{bcd}. \\
When first alternative matches \extext{d} then we know that second alternative will not match. Last choice could pass state to inform first choice about this condition.

An implementation of \verb$Switch$ is the following. We hide technical details to merge method. For details see our implementation \cite{implementation}. 

\begin{prolbox}
\begin{verbatim}
  match_memo(e,a) = 
  if memo[e,a]; memo[e,a]
  else        ; memo[e,a] <- match(e,a)

  match(Switch[ head alt merge ],a) =
    r  <- match_memo(head,a)
    r2 <- match_memo(alt[r.state],a)
    merge(r,r2)
\end{verbatim}
\end{prolbox}

\subsection{Iteration}

We use low-level repeat-until operator to represent iteration.

\noindent\begin{tabular}{| l | l | l |}
\hline
\verb$e**  $   &\verb$Many[stop e]$         & repeat-until     \\
\verb$Stop $   &\verb$Stop[stop]$           & stop operator    \\
\hline
\end{tabular}

Repeat-until can terminate if and only if we encountered corresponding \verb$Stop$ in current iteration. We add \verb$stops$ field to \verb$Args$ to collect encountered stops.

This allows to describe normal iteration \verb$e*$ and eager iteration \verb$e*?$ as \verb$(e|Stop)**$ and \verb$(Stop|e)**$ respectively.  Repeat-until is equivalent to right-recursion. For example we can flip between rules\\
\verb$ R =  a R | b      | c R | d       $\\
and\\
\verb$ R = (a   | b Stop | c   | d Stop)*$.

Except of stop condition the implementation is nearly identical to implementation of \verb$*$ operator from section \ref{simplealg}.
\begin{prolbox}
\begin{verbatim}
match(Stop[st]   ,a) = a.cont.call(  a.change(stops: a.stops+st))
match(Many[st e] ,a) = 
  cont2 <- \(a2){ 
    if a2.stops & st ; a.cont.call( a2.change(stops:a2.stops-st))
    else             ; match(e,a2.change(cont:cont2 ))
  }
  cont2.call(a)
\end{verbatim}
\end{prolbox}

\subsection{Rule call}
Rule call only affects scope of variables. When no semantic actions are present we can directly move expression to separate rule and back.

\begin{prolbox}
\begin{verbatim}
match(Rule[ e ], a) = match(e ,a)
\end{verbatim}
\end{prolbox}

For \verb$nested$ we use similar implementation as before. 

\begin{prolbox}
\begin{verbatim}
  match(Nested[st mi en],a) =
    r = match_memo(Seq[st mi en],Args[s:a.s,cont:\(m){success}])
    if r.state==success ; a.cont.call(a.change(s:r.s))
    else                ; fail
\end{verbatim}
\end{prolbox}

\section{Semantic actions} \label{semact}
A parser from section \ref{implement} is not very useful as it can only answer yes/no questions. When we integrate parser generator in programming language called {\it host language}. We can specify host language expressions called  {\it semantic actions}. For example we can use semantic actions for simple calculator:
\begin{verbatim}
  add = mul:x    '+' add:y {x+y}
      | mul
  mul = number:x '*' mul:y {x*y}
      | number
\end{verbatim}

While semantic actions are easy to add they complicate other parts of the parser.

We add the following fields:\\
\verb$Args.closure$ closure for semantic actions.\\
\verb$Args.returned$ result of last expression.\\
\verb$Result.returned$ returned result.

We model semantic act as a function that modifies arguments. For simplicity we model variable binding by semantic act.

\begin{prolbox}
\begin{verbatim}
  match( Act[ f ] ,a ) = a2 <- f.call(a.closure)
    a.cont.call(a.change(a2))
\end{verbatim}
\end{prolbox}

Now we are ready to add enter operator. 

\begin{prolbox}
\begin{verbatim}
  match( Enter[e1 e2], a) =
    match(e1,a.change(cont: \(a2){
       match(a2.change(s:a2.returned),cont:\(a3){
         a.cont.call(a3.change(s:a.s))
       }
    }
\end{verbatim}
\end{prolbox}

Semantic actions in rule invocation have shared scope. We use closure object to achieve this. A rule invocation becomes:

\begin{prolbox}
\begin{verbatim}
match( Rule[ e ] ,a ) = match(e,
                              a.change(closure:new_closure,
         cont:\(a2){ a.cont.call(a.change(s:a.s,
                                 returned:a.returned))}
     )
\end{verbatim}
\end{prolbox}

We can use {\it host language} expression called {\it semantic predicate} to decide if expression matched or not. This complicates memoization and we, for simplicity, disable memoization when semantic predicate is present.

In Amethyst we also support parametrized rules and lambdas. They are bit technical to add. For parametrized rule we first model arguments by semantic act bound to argument variables. Then we add field consisting of pairs (argument variable,parameter variable) and we initialize new closure according to pairs. For lambda we bind (expression,closure) pair to corresponding variable. We disable memoization when parametrized rule is present for same reasons as with {\it semantic predicate}.

Memoization becomes more technical. A simplest way how to get linear time complexity is to use two pass parser which in first pass run parser from section \ref{implement} and second time we just constructs parse tree. We refine this idea and run both phases in parallel. We use functor \verb$forget_semantic_actions$:

\begin{prolbox}
\begin{verbatim}
match_memo_state(e,a) =
  if (has_predicate(e) | has_predicate(a)) ; match(e,a)
  else ; e2 <- forget_semantic_actions(e)
         a2 <- forget_semantic_actions(a)
         if    memo[e2,a2]  ; memo[e2,a2]
         else               ; memo[e2,a2] <- match(e,a)

\end{verbatim}
\end{prolbox}

A simple implementation of \verb$Switch$ can be

\begin{prolbox}
\begin{verbatim}
match(Switch[ head alts merge ], a)
  r  <- match_memo_state(head,a)
  r2 <- match_memo_state(alts[r.state],a)
  if r2.state==fail
    fail(r2)
  else merge(match(head,a),
             match(alts[r.state],a))
\end{verbatim}
\end{prolbox}

Sometimes \verb$Switch$ knows that result is not needed. Then we can directly call expression simplified by \verb$forget_semantic_action$. This always happens for lookaheads.

\subsection{Time complexity}
Ford \cite{ford} rewrites iteration to recursion for linear time complexity. However most implementations naively use a loop. 

It is possible to construct test cases where arbitrary (say k) number of loops are nested together and each fails at the end of input.  This leads to
time complexity at least $n^k$ for arbitrary k. This can be seen on the following expression:\\
\begin{verbatim}
( ( ( ( 'a' )* 'b'
    / 'a' )* 'c'
  / 'a' )* 'd'
/ 'a' )* 'e'
\end{verbatim}
on \extext{aaaaaaaaaaaaaaaaaaaaaaa...}\\

We memoize continuations precisely for this reason.

For parser from section \ref{implement} there are only finitely many expressions and continuations. Thus there are only $O(n)$ memoization pairs \verb$(e,a)$.

With semantic actions we sometimes need to recalculate the result. For a given pair (\verb$nested,position$) we need to recalculate result of every \verb$(e,a)$ pair at most once.
For general \regreg{} expressions time complexity $O(n^2)$ follows. 

For structured grammars this behavior cannot happen. We do not have to recalculate when result state is \verb$fail$ or we match in lookaheads. What is left is that we could have two invocations of same \verb$nested$ expression with two different positions that recalculates same \verb$(e,a)$ pair. But this would mean that both invocations will be accepted with same end position which is in contradiction with condition 2.2 of structured grammars. Consequently the parser of structured grammars runs in linear time

With semantic predicates we can not give any complexity guarantee. To integrate them correctly we disable memoization when continuation contains semantic predicate.
\section{Memory consumption of  \regreg{} parsers}
Mizushima et al  \cite{memo} propose way to decrease the memory usage. We describe similar but simpler approach. 

The parser implementation maintain set of live branches in a list \verb$live$. The list is maintained in the following way:

\begin{itemize}
\item When parser descends into choice operator then its branches are added to \verb$live$ list.
\item When parser descends into branch, then it is removed from \verb$live$ list.
\item When parser encounters cut then branches that were cut are removed from \verb$live$ list.
\end{itemize}

When \verb$live$ list is empty we know that subsequent parsing cannot return to position smaller than current. We can safely delete all \verb$memo$ entries with smaller position.

One can observe that \verb$live$ list is not really needed.  The implementation can be further simplified by only keeping
track of the size of the list in a counter \verb$alternatives$.

The parser then deletes stale entries from memo table lazily.  It keeps
track of the rightmost position where \verb$alternatives$ was zero.  At a time
table expansion is needed, all earlier entries are deleted. This avoids the
need for the expansion if the table after deletion is at most half full.

Note that if we want to incorporate destructive semantic actions we can in same way defer their evaluation until \verb$alternatives$ is zero.

For practical grammars this extension gives nearly constant memory usage. However we can construct examples where this approach does not help, for example in expression:\\
\verb$ exp* 'x' | exp* 'y' $ \\
we need to keep memoization entries until end is reached.

\section{From \regreg{} back to $\mathrm{REG}$}
We establish a $reg$ functor. This notion gives unified way to analyze \regreg{} expressions. 

A $reg$ functor assigns to each relativized regular expression $e$ an regular expression $reg(e)$. A $reg(e)$ satisfies approximation condition that if $e$ accepts $s$ then $reg(e)$ accepts $s$ but converse is not necessary true.

We can extract useful information testing if the intersection with an suitable regular language is empty. 

\noindent\verb$  empty(e)           = reg(e    )$ $\mathtt{\cap}$\verb$ reg( ''   )$\\
         \verb$  first_char('c',e)  = reg(e    )$ $\mathtt{\cap}$\verb$ reg('c' .*)$\\
         \verb$  overlap(e1,e2)     = reg(e1 .*)$ $\mathtt{\cap}$\verb$ reg(e2  .*)$\\

If \verb$overlap(e1,e2)$ does not match anything then we can freely flip between \verb$e1|e2$ and \verb$e2|e1$. Also note that if this occurs a choice is deterministic and we do not have to backtrack if first alternative happens.

Mizushima \cite{memo} also transforms grammar to more deterministic one. We use stronger analysis. Using \verb$overlap$ we can determine where we can insert return states that inform \verb$Switch$ that next alternatives can not occur\footnote{We can also consider continuations for better results}. This transformation is quite technical and beyond scope of this paper.

While bounds \verb$minsize(e)$,\verb$maxsize(e)$ on minimal and maximal sizes of string that matches \verb$e$ can be discovered by intersecting with suitable languages it is faster to compute them directly by depth first search.

Functor $reg$ can be defined in the following way:\\
\verb$reg( 'c'                  ) = c$\\
\verb$reg( r                    ) = reg(r)$\\
\verb$reg( a* )                   = reg(a)*$\\
\verb$reg( nested(start,mid,end)) = reg(start) .* reg(end)$\\
\verb$reg( a   b )                = reg(a)   reg(b)$\\
\verb$reg( a | b )                = reg(a) | reg(b)$\\
\verb$reg(&a   b )                = reg(a) $ $\mathtt{\cap}$\verb$ reg(b)$

We use rough approximation of middle of \verb$nested$. In typical case inside nesting could be practically anything so trying to improve this approximation leads only to larger expressions without any new insights.

We shall remark that better result can be obtained by first using relativized regular machine and then converting to regular machine. This gives two advantages.

First is that \verb$Switch$ describes also lookaheads and we can describe intersection by lookahead.

Second is that we can use facts:\\
If \verb$A$ is unambiguous then \texttt{A B $\cap$ A C = A ( B $\cap$ C)}.\\
If \verb$A$ is unambiguous then \texttt{A B | A C = A ( B | C)}.

As there only finitely many $(continuation,cuts,stops)$ triples size of our machine is finite.

Note that we can test emptiness more effectively when we construct finite state machines lazily.

We do not include optimizations using $reg$ functor in this paper but in separate technical report \cite{transform}.
\section{Problems of left recursion}\label{leftrec}
Left recursion handling deserves topic of its own. Various approaches were suggested and various counterexamples found.

In \peg{}  implementing left recursion correctly is an impossible task. Consider rule:

\begin{verbatim}
L = &( L 'cd'   ) 'abc' # a -> abc -> abcbc -> ab
  | &( L 'bcd'  ) 'ab'  #       ^               |
  | L 'bc'              #       |               V
  | L 'cb'              #     abcbcb     <-   abcb
  | 'a'
\end{verbatim}

On \extext{abcbcbcd}. 

It creates infinite cycle in the recursion. This problem is more fundamental as there is an paradox:

\begin{verbatim}
  L = ~L
\end{verbatim}

We reject such self references and raise a error when lookahead refers to possibly indirectly left recursive rule. Note that in boolean grammars same problem was recognized \cite{bool}.

Left recursion can be handled by recursive descend/ascend. A rule:\\
\verb$ L = L 'bc' | L 'c' | 'ab' | 'a'$ \\
on \extext{abc}
is recognized as \extext{(a(bc))} by recursive descend parser but as \extext{((ab)c)} by recursive ascend one.
All previous approaches in \peg{} and context-free bottom-up parser used a recursive ascend variant of left recursion.
A simplest algorithm is attributed to Paull \cite{aho}. It consist of rewriting direct left recursion to equivalent rule: \\
\verb$L = L a | b | L c | d        $ \\
\verb$L =      (b |       d) (a|c)*$. 

An indirect left recursion is removed by inlining and thus reducing to direct recursion case.

In 1965 Kuno \cite{kuno} suggested to limit recursion depth by $n$. It was rejected in \peg{} setting as in presence of semantic predicates some recursive rules need more than $n$ calls. Also it was not clear how handle infinite streams. But it was rejected prematurely. 

Using $reg$ functor (or simple dataflow) we can for each expression compute lower bound on minimal length of a string that matches that expression. Using this information we can easily estimate minimal size of current continuation. When this bound exceeds the length of our string we can fail.

 For infinite streams we can guess bound by guessing initially 1 and doubling bound when recursion could continue.  We do not use this approach as it has an exponential complexity in worst case. 

Note that same technique can improve to Frost's algorithm \cite{frost}.

In packrat setting Ford used Paull algorithm to remove direct left recursion. He rejected to support left recursion with the following reason \cite{ford}: 

``At least until left recursion in TDPL is studied further, utilizing such a feature would amount
to opening a syntactic Pandora's Box, which clearly defeats the pragmatic purpose for which
the simple left recursion transformation is provided.''

Warth, Douglass, Millstein \cite{warth} attempted to add runtime detection of left recursion. With bit of imagination it could be interpreted as doing Paull algorithm at runtime. However this approach has several flaws.

One discovered by Tratt \cite{tratt} is that seed growing introduces ambiguity of direct left recursion when right recursive alternative is also present.

A revised algorithm of Tratt still contains a flaw. Tratt at certain times forbids expansion of right recursion.

Tratt approach fails to handle right-recursive lookahead as following counterexample shows.

\begin{verbatim}
  L = L 'a'
    | ~('b' L) 'b'
    | 'c'
\end{verbatim}

Third issue was discovered by Peter Goodman \cite{goodman}. Warth algorithm does not handle following grammar.

\begin{verbatim}
A = A 'a'  / B
B = B 'b'  / A  / C
C = C 'c'  / B  / 'd'
\end{verbatim}

Medeiros in unpublished paper \cite{medeir} devised a revised version of seed growing algorithm.

One of possible advantages of seed growing could be support of higher order parametrized rules. Authors Amethyst parser can in practical setting resolve higher order functions making this point a moot one.

\subsection{Left recursion in \regreg{} parser}

We combine two techniques. First we just rewrite recursion by Paull algorithm. A second technique is that continuation passing style does implicit finite state machine minimization. This is simpler and leads to smaller grammars than Moore's left corner transform heuristic \cite{moore}.

We handle left recursion inside iteration by unrolling one level. 

With some bookkeeping we can transform left recursion to recursive descend. Idea is that each alternative returns its derivation an we choose a lexicographically smallest in recursion tree. This can be done in $O(1)$ time using dynamic lowest common ancestor \cite{lca}.
\section{Summary}
We introduced notion of structured grammars that allow generation of practical linear time parsers. Our \regreg{} class generalizes \peg{} and is more suitable for various optimizations. We wrote $C$ implementation as proof of concept.

A $reg$ functor gives us unified framework for many optimizations. 

We also integrated left recursion into our parser. 

Author also developed a dynamic parser of structured grammars \cite{dynami}. The dynamic parser allows to modify string and ask for updated parse results Our dynamic parser recomputes only rules that it must recompute. For practical grammars overhead is within constant factor of time spend on recomputing rules. On worst case we multiply time  spend on recomputing rules by logarithmic factor.

Structured grammar are promising for integration with IDE. With information that structured grammars expose we can automatically offer code folding, error reporting and syntax highlighter \cite{example}.

Author developed structured grammars as part of Amethyst language. Amethyst generalizes pattern matching in several directions as is described in authors thesis \cite{thesis}.


\begin{thebibliography}{99}
\bibitem{aho}
V. Aho, R. Sethi, and J. D. Ullman.
\emph{Compilers: Principles, Techniques, and Tools.}
Addison-Wesley Publishing Company, Reading,
Massachusetts. 1986.
\bibitem{thesis} \myname, \emph{Pattern matching in compilers}, Master's thesis, in preparation.
\bibitem{dynami} \myname, \emph{Dynamic top-down parsing}, in preparation.
\bibitem{transform} \myname, \emph{Transformations of \regreg{} expressions}, in preparation.
\bibitem{example} \verb$http://kam.mff.cuni.cz/~ondra/amethyst/parser_highlight.ame.html$.
\bibitem{implementation} \verb$http://kam.mff.cuni.cz/~ondra/regreg$.
\bibitem{lca} Richard Cole,  Ramesh Hariharan \emph{Dynamic LCA queries on trees,} Proceeding SODA '99 Proceedings of the tenth annual ACM-SIAM symposium on Discrete algorithms
\bibitem{ford} Ford Bryan, \emph{Packrat Parsing: a Practical Linear-Time Algorithm with Backtracking}. Master's thesis, MIT, September 2002
\bibitem{frost}Frost, R.; R. Hafiz and P. Callaghan (June 2007).\emph{ Modular and Efficient Top-Down Parsing for Ambiguous Left-Recursive Grammars.}. 10th International Workshop on Parsing Technologies (IWPT), ACL-SIGPARSE (Prague)
\bibitem{goodman} Goodman Peter,  \emph{ Re: [PEG] Res: Problem w/ nullable left recursion and trailing rules in "Packrat Parsers Can Support Left Recursion"}
\verb$http://www.mail-archive.com/peg@lists.csail.mit.edu/msg00185.html$, 2008
\bibitem{john} M. Johnson. \emph{Memoization in top-down parsing.} Comput. Linguist., 21(3): 405-417, 1995.
\bibitem{kuno} Kuno, S. \emph{The predictive analyzer and a path elimination technique.} Comm. ACM 8(7) 453–462, 1965.
\bibitem{oreg}  Sergio Medeiros, Fabio Mascarenhas, and Roberto Ierusalimschy, \emph{From Regular Expressions to Parsing Expression Grammars.}  SBLP, September 2011.
\bibitem{medeir} Sergio Medeiros,\emph{ Left Recursion in PEGs} \\ \verb$http://www.lua.inf.puc-rio.br/~sergio/leftpeglist.pdf$, 2010
\bibitem{memo} Mizushima, K., Maeda, A., Yamaguchi, Y.: \emph{Packrat parsers can handle practical grammars in mostly
constant space}, Proceedings of the 9th ACM SIGPLAN-SIGSOFT Workshop on Program Analysis for
Software Tools and Engineering, PASTE'10, Toronto, Ontario, Canada, June 5-6, 2010 (S. Lerner,
A. Rountev, Eds.), ACM, 2010.
\bibitem{moore} C. Moore. \emph{Removing left recursion from context-free grammars.} In
Proc. 1st North American chapter of the Association for Computational Linguistics conference, pages 249–255, 2000.
\bibitem{bool} Okhotin Alexander,  \emph{LR parsing for Boolean grammars}, International Journal of Foundations of Computer Science 17:3 (2006), 629--664.
\bibitem{nfa} Rabin, M. O. and Scott, D. \emph{Finite automata and their decision problems} IBM J. Res. Dev. vol 3, 1959
\bibitem{radz1} Roman Redziejowsky, BITES instead of FIRST for Parsing Expression Grammar Fundamenta Informaticae 109, 3 (2011) 323-337.
\bibitem{katahdin} Christopher Graham Seaton, \emph{A Programming Language Where the Syntax and Semantics Are Mutable at Runtime}, University of Bristol, Master thesis, \verb$http://www.chrisseaton.com/katahdin/katahdin.pdf$
\bibitem{tomita} Tomita, Masaru \emph{An efficient augmented-context-free parsing algorithm},Comput. Linguist. 13, 31--46, 1987
\bibitem{tratt} Laurence Tratt, \emph{Direct Left-recursive Parsing Expression Grammars.} Technical report EIS-10-01, Middlesex University, October 2010
\bibitem{valiant} Valiant, Leslie G. \emph{General context-free recognition in less than cubic time,} Journal of Computer and System Sciences 10, 1975
\bibitem{warth} Alessandro Warth, James R. Douglass, and Todd Millstein, \emph{ Packrat parsers can support left recursion} PEPM '08, 1998\end{thebibliography}
\end{document}